\documentclass[a4paper,10pt,3p,final,pdftex]{elsarticle}

\biboptions{sort&compress}

\usepackage{amsmath}
\usepackage{amsfonts}
\usepackage{graphicx}
\usepackage{algorithm,algorithmic}
\usepackage{url}
\usepackage{color}
\usepackage{tikz}
\usepackage{multirow}

\newdefinition{remark}{Remark}
\newcommand {\RM}[1]{\mathrm{#1}}
\renewcommand{\Re}{\mathbb{R}}

\newcommand{\opal}{\textsc{OPAL}}
\newcommand{\ippl}{\textsc{IP$^2$L}}
\newcommand{\oursolver}{\textsc{SAAMG-PCG}}

\journal{Journal of Computational Physics}

\begin{document}

\begin{frontmatter}

\title{A Fast Parallel Poisson Solver on Irregular Domains Applied to Beam
  Dynamic Simulations}

\author[psi]{A.~Adelmann\corref{cor}}
\ead{andreas.adelmann@psi.ch}

\author[eth]{P.~Arbenz}
\ead{arbenz@inf.ethz.ch}

\author[psi,eth]{Y.~Ineichen}
\ead{ineichen@inf.ethz.ch}

\address[psi]{Paul Scherrer Institut, CH-5234 Villigen, Switzerland}
\address[eth]{ETH Z\"urich, Chair of Computational Science,
  Universit\"atsstrasse 6, CH-8092 Z\"urich, Switzerland}

\cortext[cor]{Corresponding author}

\begin{abstract}
  We discuss the scalable parallel solution of the Poisson equation
  within a Particle-In-Cell (PIC) code for the simulation of electron
  beams in particle accelerators of irregular shape.  The problem is
  discretized by Finite Differences.  Depending on the treatment of the
  Dirichlet boundary the resulting system of equations is symmetric or
  `mildly' nonsymmetric positive definite.  In all cases, the system is
  solved by the preconditioned conjugate gradient algorithm with
  smoothed aggregation (SA) based algebraic multigrid (AMG)
  preconditioning.  We investigate variants of the implementation of
  SA-AMG that lead to considerable improvements in the execution times.
  We demonstrate good scalability of the solver on distributed memory
  parallel processor with up to 2048 processors.  We also compare our
  \oursolver\ solver with an FFT-based solver that is more commonly used
  for applications in beam dynamics.
\end{abstract}


\end{frontmatter}

\nocite{*}
 
\section{Introduction}
\label{sec:intro}

In recent years, precise beam dynamics simulations in the design of
high-current low-energy hadron machines as well as of 4th generation
light sources have become a very important research topic.  Hadron
machines are characterized by high currents and hence require excellent
control of beam losses and/or keeping the emittance (a measure of the
phase space) of the beam in narrow ranges. This is a challenging problem
which requires the accurate modeling of the dynamics of a large ensemble
of macro or real particles subject to complicated external focusing,
accelerating and wake-fields, as well as the self-fields caused by the
Coulomb interaction of the particles.  In general the geometries of
particle accelerators are large and complicated. The discretization of
the computational domain is time dependent due to the relativistic
nature of the problem.  time and space dilatation. Both phenomenas have
a direct impact on the numerical solution method.  
 
The solver described in this paper is part of a general accelerator
modeling tool \opal~(Object Oriented Parallel Accelerator Library)
\cite{opal}.  \opal\ allows to tackle the most challenging problems in
the field of high precision particle accelerator modeling.  These
include the simulation of high power hadron accelerators and of next
generation light sources.

Some of the effects can be studied by using a low dimensional model,
i.e., envelope equations~\cite{sach:68, sach:71, stru-reis:1984,
  gluckstern1}.  These are a set of ordinary differential equations for
the second-order moments of a time-dependent particle distribution.
They can be calculated fast, however the level of detail is mostly 
insufficient for quantitative studies.  Furthermore, a priori knowledge of
critical beam parameters such as the emittance is required with the consequence that the
envelope equations cannot be used as a self-consistent method.

One way to overcome these limitations is by considering the
Vlasov-Poisson description of the phase space, including external and
self-fields and, if needed, other effects such as wakes.  To that end
let $f(\mathbf{x},\mathbf{v},t)$ be the density of the particles in the
phase space, i.e., the position-velocity $(\mathbf{x}, \mathbf{v})$
space.  Its evolution is determined by the collisionless \emph{Vlasov
  equation},
\begin{equation}\label{eq:Vlasov}
  \frac{df}{dt}=\partial_t f + \mathbf{v} \cdot \nabla_{\mathbf{x}} f
  +\frac{q}{m_0}(\mathbf{E}+ \mathbf{v}\times\mathbf{B})\cdot
  \nabla_{\mathbf{v}} f  =  0,
\end{equation}
where $m_0$, $q$ denote particle mass and charge, respectively.  The
electric and magnetic fields $\mathbf{E}$ and $\mathbf{B}$ are
superpositions of external fields and self-fields (space charge),
\begin{equation}\label{eq:allfield}
    \mathbf{E} =
    \mathbf{E_{\RM{ext}}} + \mathbf{E_{\RM{self}}}, \quad
    \mathbf{B} =
    \mathbf{B_{\RM{ext}}} + \mathbf{B_{\RM{self}}}.
\end{equation}
If $\mathbf{E}$ and $\mathbf{B}$ are known, then each particle can be
propagated according to the equation of motion for charged particles in an
electromagnetic field,
\begin{equation*}\label{eq:motion}
  \frac{d\mathbf{x}(t)}{dt}  = \mathbf{v},
  \quad
  \frac{d\mathbf{v}(t)}{dt}  = \frac{q}{m_0}\left(\mathbf{E} +
    \mathbf{v}\times \mathbf{B}\right).
\end{equation*}

After the movement of the particles 
$\mathbf{E_{\RM{self}}}$ and $\mathbf{B_{\RM{self}}}$ have to be updated.  
To that end we change the coordinate system into the one moving with the
particles.  By means of the appropriate \emph{Lorentz
  transformation}~\cite{lali:84} we arrive at a (quasi-) static
approximation of the system in which the transformed magnetic field
becomes negligible, $\hat{\mathbf{B}}\! \approx\! \mathbf{0}$.  The
transformed electric field is obtained from
\begin{equation}\label{eq:e-field}
  \hat{\mathbf{E}}=\hat{\mathbf{E}}_{\RM{self}}=-\nabla\hat{\phi},
\end{equation}
where the electrostatic potential $\hat{\phi}$ is the solution of the
\emph{Poisson problem}
\begin{equation}\label{eq:poisson0}
  - \Delta \hat{\phi}(\mathbf{x}) =
  \frac{\hat{\rho}(\mathbf{x})}{\varepsilon_0},
\end{equation}
equipped with appropriate boundary conditions, see
section~\ref{sec:discr}.  Here, $\hat{\rho}$ denotes the spatial charge
density and $\varepsilon_0$ is the dielectric constant.
By means of the inverse Lorentz transformation the electric field
$\hat{\mathbf{E}}$ can then be transformed back to yield both the
electric and the magnetic fields in~\eqref{eq:allfield}.

The Poisson problem~\eqref{eq:poisson0} discretized by finite
differences can efficiently be solved on a rectangular grid by a
Particle-In-Cell (PIC) approach~\cite{qiry:01}.  The right hand side
in~\eqref{eq:poisson0} is discretized by sampling the particles at the
grid points.  In~\eqref{eq:e-field}, $\hat{\phi}$ is interpolated at the
particle positions from its values at the grid points. We also note that
the FFT-based Poisson solvers and similar
approaches~\cite{qiry:01,qigl:04} are restricted to box-shaped or open domains.

Serafini et al.~\cite{serafini_2005} report on a state-of-the-art
conventional FFT-based algorithm for solving the Poisson equation with
`infinite-domain', i.e., open boundary conditions for large problems in
accelerator modeling.  The authors show improvements in both accuracy
and performance, by combining several techniques: the method of local
corrections, the James algorithm, and adaptive mesh refinement.

However with the quest of high intensity, high brightness beams together
with ultra low particle losses, there is a high demand to consider the
true geometry of the beam-pipe in the numerical model. This assures that
the image charge components are taken properly into account. This
results in a more exact modeling of the non-linear beam dynamics which
is indispensable for the next generation of particle accelerators.

In a related paper by P{\"o}plau et
al.~\cite{poplau_self-adaptive_2008}, an iterative solver preconditioned
by geometric multigrid is used to calculate space-charge forces.  The
authors employ a mesh with adaptive spacings to reduce the workload of
the BiCGStab solver used to solve the nonsymmetric system arising from
quadratic extrapolation at the boundary.  The geometric multigrid solver
used in their approach is much more sensitive to anisotropic grids
arising in beam dynamic simulations (e.g.\ special coarsening operators
have to be defined).  With smoothed aggregation-based algebraic multigrid
(AMG) preconditioning as used in this paper the aggregation smoother takes
care of anisotropies and related issues and leads to a robustness superior
to geometric multigrid, see~\cite{trcl:09} for a discussion.  The
preconditioner easily adapts to the elongation  of the computational domain
that happens during our simulation.

In Section~\ref{sec:discr} we describe how the Poisson equation on a
`general' domain $\Omega \subset \Re^3$ can be solved by finite
differences and the PIC approach.  We treat the boundary in three
different ways, by constant, by linear, and by quadratic extrapolation,
the latter being similar to the approach of McCorquodale \textit{et
  al.}~\cite{mcgv:04}.  The system of equation is solved by the
conjugate gradient algorithm preconditioned by smoothed aggregation
AMG~\cite{vamb:96a, tuto:00}, see Section~\ref{sec:method}. For this solver we coined the acronym: \oursolver.  The preconditioned conjugate
gradient~(PCG) algorithm is also used if the system is `mildly'
nonsymmetric.  In Section~\ref{sec:impl} we deal with details of the
implementation, in particular its parallelization.  In
Section~\ref{sec:results} we report on numerical experiments including a
physical application from beam dynamics.  In Section~\ref{sec:concl} we
draw our conclusions.


\section{The discretization}
\label{sec:discr}

In this section we discuss the solution of the Poisson equation in a
domain 
\begin{figure}[htb]
  \centering
  \includegraphics[width=0.42\textwidth]{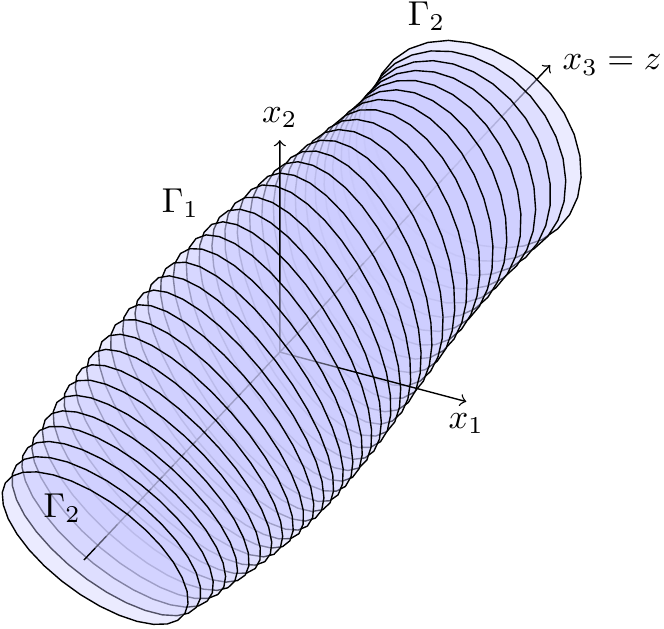}
  \caption{Sketch of a typical domain}
  \label{fig:domain}
\end{figure}
$\Omega \subset \Re^3$ as indicated in Figure~\ref{fig:domain}.  The
boundary of the domain is composed of two parts, a curved, smooth
surface $\Gamma_1$ and two planar portions at $z=-d$ and $z=+d$ that
form together $\Gamma_2$.  In physical terms $\Gamma_1$ forms the casing
of the pipe, while $\Gamma_2$ is the open boundary at the inlet and
outlet of the beam pipe, respectively.  The centroid of the particle bunch is at the
origin of the coordinate system.  The Poisson problem that we are going
to solve is given by
\begin{equation} \label{eq:poisson}
  \begin{aligned}
    -\Delta \phi &= \frac{\rho}{\epsilon_0}\ \text{in}\ \Omega, \\
    \phi &= g \equiv 0\ \text{on}\ \Gamma_1,   \\
    \frac{\partial \phi}{\partial \mathbf{n}} + \frac {1}{d} \phi &= 0\
    \text{on}\ \Gamma_2.
  \end{aligned}
\end{equation}
The parameter $d$ in the Robin boundary condition denotes the distance
of the charged particles to the boundary
\cite{poplau_self-adaptive_2008}.  It is half the extent of $\Omega$ in
$z$-direction.  Notice that the Robin boundary condition applies only on
the planar paraxial portions of the boundary.

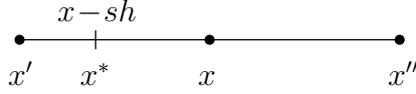
\begin{figure}[htb]
  \centering
\begin{picture}(0,0)%
\includegraphics{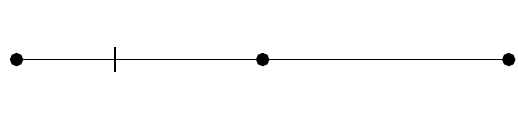}%
\end{picture}%
\setlength{\unitlength}{2072sp}%
\begingroup\makeatletter\ifx\SetFigFontNFSS\undefined%
\gdef\SetFigFontNFSS#1#2#3#4#5{%
  \reset@font\fontsize{#1}{#2pt}%
  \fontfamily{#3}\fontseries{#4}\fontshape{#5}%
  \selectfont}%
\fi\endgroup%
\begin{picture}(4718,1191)(2101,-4324)
\put(4366,-4201){\makebox(0,0)[lb]{\smash{{\SetFigFontNFSS{12}{14.4}{\familydefault}{\mddefault}{\updefault}$x$}}}}
\put(6616,-4201){\makebox(0,0)[lb]{\smash{{\SetFigFontNFSS{12}{14.4}{\familydefault}{\mddefault}{\updefault}$x''$}}}}
\put(2116,-4201){\makebox(0,0)[lb]{\smash{{\SetFigFontNFSS{12}{14.4}{\familydefault}{\mddefault}{\updefault}$x'$}}}}
\put(2701,-3436){\makebox(0,0)[lb]{\smash{{\SetFigFontNFSS{12}{14.4}{\familydefault}{\mddefault}{\updefault}$x\!-\!sh$}}}}
\put(2971,-4201){\makebox(0,0)[lb]{\smash{{\SetFigFontNFSS{12}{14.4}{\familydefault}{\mddefault}{\updefault}$x^*$}}}}
\end{picture}%
  \caption{1-dimensional sketch of a near-boundary point $x$}
  \label{fig:boundary}
\end{figure}

We discretize~\eqref{eq:poisson} by a second order finite difference
scheme defined on a rectangular lattice (grid)
\begin{displaymath}
  \Omega_h:=\left\{ \mathbf{x} \in {\Omega}\cup\Gamma_2 \ |\ x_i/h_i \in
    \mathbb{Z} \ \mbox{for}\ i=1,2,3 \right\},
\end{displaymath}
where $h_i$ is the grid spacing and $\mathbf{e}_i$ the unit vector in the $i$-th coordinate direction. The
grid is arranged in a way that the two portions of $\Gamma_2$ lie in
grid planes.  A lattice point is called an \emph{interior} point if all
its direct neighbours are in $\Omega$.  All other grid points are called
\emph{near-boundary} points.  At interior points $\mathbf{x}$ we
approximate $\Delta u (\mathbf{x})$ by the well-known 7-point difference
star
\begin{equation}  \label{eq:7pt-star}
  -\Delta_h u(\mathbf{x}) = 
  \sum_{i=1}^3
  \frac{-u(\mathbf{x}\!-\!h_i\mathbf{e}_i) + 2 u(\mathbf{x})
  - u(\mathbf{x}\!+\!h_i\mathbf{e}_i)}{h_i^2}.
\end{equation}
At grid points near the boundary we have to take the boundary conditions
in~\eqref{eq:poisson} into account.  To explain the schemes on the
Dirichlet (or PEC) boundary $\Gamma_1$ let $\mathbf{x}$ be a
near-boundary point.  Let $\mathbf{x}' := \mathbf{x} - h_i\mathbf{e}_i$
for some $i$ be outside $\Omega$ and let $\mathbf{x}^* := \mathbf{x} - s
h_i\mathbf{e}_i$, $0<s\le1$, be the boundary point between $\mathbf{x}$
and $\mathbf{x}'$ that is closest to $\mathbf{x}$,
cf.~Figure~\ref{fig:boundary}.
If $s=1$, i.e., if $\mathbf{x}'\in \partial\Omega$ then $u(\mathbf{x}')$
in~\eqref{eq:7pt-star} is replaced by the prescribed boundary value.
Otherwise, we proceed in one of the three following ways~\cite{fowa:60,
  hack:94}
\begin{enumerate}
\item In \emph{constant extrapolation} the boundary value prescribed at
  $\mathbf{x} - s h_i\mathbf{e}_i \in \Gamma_1$ is assigned to
  $u(\mathbf{x}')$,
  \begin{equation}
    \label{eq:const_extrapol}
    u(\mathbf{x}') = u(\mathbf{x} - h_i\mathbf{e}_i) := g(\mathbf{x}^*).
  \end{equation}
\item In \emph{linear extrapolation} the value at $\mathbf{x}'$ is
  obtained by means of the values $u$ at $\mathbf{x}$ and at $\mathbf{x}
  - sh_i\mathbf{e}_i$,
  \begin{equation}
    \label{eq:lin_extrapol}
    u(\mathbf{x}') := (1-\frac{1}{s})\, u(\mathbf{x}) + \frac{1}{s}\,
    g(\mathbf{x}^*).
  \end{equation}


\item \emph{Quadratic extrapolation} amounts to the Shortley-Weller
  approximation~\cite{fowa:60,shwe:39,mcgv:04}.  If $\mathbf{x}'' :=
  \mathbf{x} + h_i\mathbf{e}_i \in \Omega_h$ then the value
  $u(\mathbf{x}')$ is obtained by quadratic interpolation of the values
  of $u$ at $\mathbf{x}$, $\mathbf{x}''$, and the boundary point
  $\mathbf{x}^*$,
  \begin{equation} \label{eq:quad_extrapol}
    u(\mathbf{x}') := \frac{2(s\!-\!1)}{s} u(\mathbf{x}) -
    \frac{s\!-\!1}{s\!+\!1} u(\mathbf{x''}) + 
    \frac{2}{s(s\!+\!1)} g(\mathbf{x}^*).
  \end{equation}
  If $\mathbf{x}'' \not\in \Omega_h$ then let $\mathbf{x}^{**} :=
  \mathbf{x} + t h_i\mathbf{e}_i$, $0<t\le1$, be the boundary point
  between $\mathbf{x}$ and $\mathbf{x}''$ that is closest to
  $\mathbf{x}$.  Then, similarly as before, we get
  \begin{equation} \label{eq:quad_extrapol1}
    \begin{aligned}
      u(\mathbf{x}') &:= \frac{(s\!-\!1)(t\!+\!1)}{st} u(\mathbf{x}) +
      \frac{(t\!+\!1)}{(s\!+\!t)s} g(\mathbf{x}^*) -
      \frac{(s\!-\!1)}{(s\!+\!t)t} g(\mathbf{x}^{**}), \\[1mm]
      u(\mathbf{x}'') &:= \frac{(s\!+\!1)(t\!-\!1)}{st} u(\mathbf{x}) -
      \frac{(t\!-\!1)}{(s\!+\!t)s} g(\mathbf{x}^*) +
      \frac{(s\!+\!1)}{(s\!+\!t)t} g(\mathbf{x}^{**}).
    \end{aligned}
  \end{equation}
 
\end{enumerate}
In all extrapolation formulae given above we set $g(\mathbf{x}^*) =
g(\mathbf{x}^{**}) = 0$ according to~\eqref{eq:poisson}.  The value on
the right side of~\eqref{eq:const_extrapol}--\eqref{eq:quad_extrapol1}
substitutes $u(\mathbf{x}\! \pm \! h_i\mathbf{e}_i)$
in~\eqref{eq:7pt-star}.

Let us now look at a grid point $\mathbf{x}$ on the open boundary
$\Gamma_2$.  If $\mathbf{x}$ is located on the inlet of the beam pipe
then $\mathbf{x}'':=\mathbf{x}\! +\! h_3\mathbf{e}_3 \in \Omega$ and
$\mathbf{x}':=\mathbf{x}\! -\! h_3\mathbf{e}_3 \not\in \Omega$.  The
Robin boundary condition is approximated by a central difference,
\begin{displaymath}
  - \frac{u(\mathbf{x}'') - u(\mathbf{x}')}{2h_3}
  + \frac{1}{d}u(\mathbf{x}) = 0,
\end{displaymath}
or
\begin{equation}  \label{eq:inlet}
  u(\mathbf{x}') = u(\mathbf{x}'') - \frac{2h_3}{d}u(\mathbf{x}).
\end{equation}
The same formula holds on the outlet boundary portion if $\mathbf{x}'$
denotes the virtual grid point outside $\Omega$.

Notice that some lattice points may be close to the boundary with regard
to more than one coordinate direction.  Then, the
procedures~\eqref{eq:const_extrapol}--\eqref{eq:inlet} must be
applied to all of them.  

The finite difference discretization just described leads to a system of
equations
\begin{equation} \label{eq:lin-syst}
  A \mathbf{x} = \mathbf{b},
\end{equation}
where $\mathbf{x}$ is the vector of unknown values of the potential and
$\mathbf{b}$ is the vector of the charge density interpolated at the grid
points.
The Poisson matrix $A$ is an $M$-matrix irrespective of the boundary
treatment~\cite{hack:94}.  Constant and linear extrapolation lead to a
\emph{symmetric} positive definite $A$ while quadratic extrapolation
yields a \emph{nonsymmetric} but still positive definite Poisson matrix.

Notice that the boundary extrapolation introduces large diagonal
elements in $A$ if $s$ or $t$ gets close to zero.  In order to avoid
numerical difficulties it is advisable to scale the system matrix.  If
$D = \mbox{diag}{A}$, then we replace $A$ in~\eqref{eq:lin-syst} by
$D^{-1/2} A D^{-1/2}$ and adapt $\mathbf{x}$ and $\mathbf{b}$
accordingly.  

\section{The solution method}
\label{sec:method}

In this section we discuss the solution of~\eqref{eq:lin-syst}, the
Poisson problem~\eqref{eq:poisson} discretized by finite differences as
described in the previous section.

\subsection{The conjugate gradient algorithm}

The matrix $A$ in~\eqref{eq:lin-syst} is symmetric positive definite
(spd) if the boundary conditions are treated by constant or linear
extrapolation.  For symmetric positive definite systems, the conjugate
gradient (CG) algorithm~\cite{hack:94,hest:52} provides a fast and
memory efficient solver.  The CG algorithm minimizes the quadratic
functional
\begin{equation} \label{eq:cg-funct}
  \varphi(\mathbf{x}) = \frac{1}{2}\mathbf{x}^T A \mathbf{x} - \mathbf{x}^T
  \mathbf{b}
\end{equation}
in the Krylov space that is implicitly constructed in the iteration.  In
the $k$-th iteration step the CG algorithm minimizes the quadratic
functional $\varphi$ along a search direction $\mathbf{d}_k$.  The search
directions turn out to by pairwise conjugate, $\mathbf{d}_k^T A
\mathbf{d}_j = 0$ for all $k\neq j$, and $\varphi(\mathbf{x})$ is
minimized in the whole $k$-dimensional Krylov space.

If we use the quadratic extrapolation~\eqref{eq:quad_extrapol} at the
boundary then $A$ in~\eqref{eq:lin-syst} is not symmetric positive
definite anymore.  Nevertheless, the solution of~\eqref{eq:lin-syst} is
still a minimizer of $\varphi(\mathbf{x})$.  The CG algorithm can be
used to solve~\eqref{eq:lin-syst}.  It is known to
converge~\cite{gree:97}.  However, the finite termination property of CG
is lost as the search directions are not mutually conjugate any more.
Only consecutive search directions are conjugate, $\mathbf{d}_k^T A
\mathbf{d}_{k-1} = 0$, reflecting the fact that $\varphi(\mathbf{x})$ is
minimized only locally.
Young \& Jea~\cite{yoje:80} investigated generalizations of the
conjugate gradient algorithm for nonsymmetric positive definite
matrices, in which conjugacy is enforced among $\mathbf{d}_k, \ldots,
\mathbf{d}_{k-s}$ for some $s>1$.  We do not pursue these approaches
here.  As GMRES~\cite{sasc:86}, they consume much more memory space than
the straightforward CG method that turned out to be extremely efficient
for our application.  Although $A$ is nonsymmetric it is so only
`mildly', i.e., there are some deviations from symmetry only at some of
the boundary points.  Therefore, one may hope that the conjugate
gradient method still performs reasonably well.  This is what we
actually did observe in our experiments.

Methods that are almost as memory efficient as CG like, e.g., the
stabilized biconjugate gradient (BiCGStab) method~\cite{vors:92} could
be used for solving~\eqref{eq:lin-syst}, also.  However, when
considering computational costs we note that BiCGStab requires two
matrix-vector products per iteration step, in contrast to CG that
requires only one.

\subsection{Preconditioning}

To improve the convergence behavior of the CG methods we
precondition~\eqref{eq:lin-syst}.  The preconditioned system has the
form
\begin{equation*}
  {M}^{-1}{A} \mathbf{x} = {M}^{-1}\mathbf{b},
\end{equation*}
where the positive definite matrix $M$ is the preconditioner.
A good choice of the preconditioner reduces the condition of the system
and thus the number of steps the iterative solver takes until
convergence~\cite{hack:94,gree:97}.  Preconditioning is inevitable for systems
originating in finite difference discretizations of 2nd order PDE's since
their condition number increases as $h^{-2}$ where $h$ is the mesh width~\cite{leve:07}.

In this paper we are concerned with multilevel preconditioners.
Multigrid or multilevel preconditioners are the most effective
preconditioners, in particular for the Poisson
problem~\cite{hack:85,tros:00}.  Multigrid methods make use of the
observation that a smooth error on a fine grid can be well approximated
on a coarser grid.  When this coarser grid is chosen to be a sufficient
factor smaller than the fine grid the resulting problem is smaller and
thus cheaper to solve.  We can continue coarsening the grid until we
arrive at a problem size that can be solved cheaply by a direct solver.
This observation suggests an algorithm similar to
Algorithm~\ref{alg:mg_algo}.

\begin{algorithm}
  \caption{Multigrid V-cycle Algorithm} \label{alg:mg_algo}
  \begin{algorithmic}[1]
    \STATE \textbf{procedure} MultiGridSolve($A_\ell$, $\mathbf{b}_\ell$, $\mathbf{x}_\ell$, $\ell$)
    
    \IF{$\ell = \mbox{maxLevel}-1$}
    \STATE DirectSolve $A_\ell \mathbf{x}_\ell = \mathbf{b}_\ell$
    \ELSE
    \STATE $\mathbf{x}_\ell$ $\leftarrow$ $S^{pre}_\ell$($A_\ell$, $\mathbf{b}_\ell$, $0$) \COMMENT{presmoothing}
    \STATE $\mathbf{r}_\ell$ $\leftarrow$ $\mathbf{b}_\ell$ - $A_\ell
    \mathbf{x}_\ell$ \COMMENT{calculate residual}
    \STATE $\mathbf{b}_{\ell+1}$ $\leftarrow$ $R_\ell \mathbf{r}_\ell$
    \COMMENT{restriction} 
    \STATE $\mathbf{v}_{\ell+1}$ $\leftarrow$ $\mathbf{0}$
    \STATE MultiGridSolve($A_{\ell+1}$, $\mathbf{b}_{\ell+1}$,
    $\mathbf{v}_{\ell+1}$, $\ell\!+\!1$) 
    \STATE $\mathbf{x}_\ell$ $\leftarrow$ $\mathbf{x}_\ell$ + $P_\ell
    \mathbf{v}_{\ell+1}$ \COMMENT{coarse grid correction} 
    \STATE $\mathbf{x}_\ell$ $\leftarrow$ $S^{post}_\ell$($A_\ell$,
    $\mathbf{b}_\ell$, $\mathbf{x}_\ell$) 
    \ENDIF
    \STATE \textbf{end procedure}
  \end{algorithmic}
\end{algorithm}

The procedure starts on the finest level ($\ell\!=\!0$) and repeatedly
coarsens the grid until the coarsest level is reached
(\texttt{maxLevel}) on which a direct solver is used to solve the
problem at hand.  On all other levels $\ell$ the algorithm starts by
presmoothing $S_\ell^{pre}$ the problem to damp high frequency
components of the error (line~5).  Subsequently the fine grid on level
$\ell$ can be restricted with the restriction operator $R_\ell$ to a
coarser grid on level $\ell+ 1$ (line~7).  This essentially
``transfers'' the low frequency components on the fine grid to high
frequency components on the coarse grid.  After the recursion has
reached the coarsest level and used the direct solver to solve the
coarse level problem the solution can be prolongated back to a finer
grid.  This is achieved with the prolongation operator $P_\ell$
(line~10).  Often a postsmoother $S_\ell^{post}$ is used to remove
artifacts caused by the prolongation operator.  Usually these operators
(for every level $\ell$) are defined in a setup phase preceding the
execution of the actual multigrid algorithm.  Lastly, $A_\ell$ denotes
the matrix of the discretized system in level $\ell$. 




The performance of multigrid methods profoundly depends on the choices and
interplay of the smoothing and restriction operators. To ensure that the
resulting preconditioner is symmetric we use the same pre- and postsmoother
$S_\ell$ and the restriction operator is chosen to be the transpose of the
prolongation operator $R_\ell = P_\ell^T$. This leaves us with two operators,
$P_\ell$ and $S_\ell$, that have to be defined for every level.


\paragraph{Prolongation Operator $P_\ell$} Aggregation based methods
cluster the fine grid unknowns to aggregates (of a specific form, size,
etc.) as representation for the unknowns on the coarse grid.  First,
each vertex of $G_\ell$, the adjacency graph of $A_\ell$, is assigned to
one of the pairwise disjoint aggregates.  Then, a tentative prolongation
operator matrix is formed where matrix rows correspond to vertices and
matrix columns to aggregates.  A matrix entry $(i,j)$ has a value of $1$
if the $i^{th}$ vertex is contained in $j^{th}$ aggregate and $0$
otherwise.  This prolongation operator basically corresponds to a
piecewise constant interpolation operation.  To improve robustness one
can additionally smooth the tentative prolongation operator. This is
normally done with a damped Jacobi smoother.  In general applying a
smoother results in better interpolation properties opposed to the
piecewise constant polynomials and improves convergence properties.
Tuminaro \& Tong~\cite{tuto:00} propose various strategies how to
parallelize this process.  The simplest strategy is to have each
processor aggregate its portion of the grid.  This method is called
``decoupled'' since the processors act independently of each other.
Usually the aggregates are formed as cubes of $3^d$ vertices in $d$
dimensions.  Since the domains under consideration are close to
rectangular the decoupled scheme seems to be an appropriate strategy.
In the interior of our domain we get regular cubes covering the
specified number of vertices.  Only a few aggregates near subgrid
interfaces and domain boundary contain fewer vertices resulting in a
non-optimal aggregate size.  The overhead introduced is small. 
``Coupled'' coarsening strategies, e.g. Parmetis, introduce interprocessor
communication and are often needed in the presence of highly irregular
domains.  In our context applying uncoupled methods only restrict the size
of the coarsest problem.  This is due to the fact that on the coarsest
level each processor must at least hold one degree of freedom.

\paragraph{Smoothing operator $S_\ell$} As advised in~\cite{abht:03} we
choose a Chebyshev polynomial smoother.  The choice is motivated by the
observation that polynomial smoothers perform better in parallel than
Gauss-Seidel smoothers.  Advantages are, e.g., that polynomial smoothers
do not need special matrix kernels and formats for optimal performance
and, generally, polynomial methods can profit of architecture optimized
matrix vector products.  Nevertheless, routines are needed that yield
bounds for the spectrum.  But these are needed by the prolongator
smoother anyway.

\paragraph{Coarse level solver} The employed coarse level solver
(Amesos-KLU) ships the coarse level problem to node~0 and solves it
there by means of an LU factorization.  Once the solution has been
calculated it is broadcast to all nodes.  To gather and scatter data a
substantial amount of communication is required.  Moreover the actual
solve can be expensive if the matrix possesses a large amount of
nonzeros per row.

An alternative is to apply a few steps of an iterative solver (e.g.
Gauss-Seidel) at the coarsest level.  A small number of iteration steps
decreases the quality of the preconditioner and thus increases the PCG
iteration count.  A large number of iteration steps increases the time
for applying the AMG preconditioner.  We found three Gauss-Seidel
iteration steps to be a good choice for our application.

\paragraph{Cycling method} We observed a tendency that timings for the
W-cycle are 10\% -- 20\% slower compared with the V-cycle.

\section{Implementation details}
\label{sec:impl}

The multigrid preconditioner and iterative solver are implemented with
the help of the Trilinos framework~\cite{Trilinos-Web-Site,
  Trilinos-TOMS}.  Trilinos provides state-of-the-art tools for
numerical computation in various packages.  Aztec, e.g., provides
iterative solvers and ML~\cite{gsht:06} multilevel preconditioners.  By
means of ML, we created our smoothed aggregation-based AMG
preconditioner.  The essential parameters of the preconditioner
discussed above are listed in Table~\ref{tab:sa_setup}.
\begin{table}[htb]
  \begin{center}
    \begin{tabular}{l|l}
      \hline
      name & value \\
      \hline
      preconditioner type & MGV \\
      smoother & pre and post \\
      smoother type & Chebyshev \\
      aggregation type & Uncoupled \\
      coarse level solver & Amesos-KLU \\
      maximal coarse level size & 1000 \\
      \hline
    \end{tabular}
    \caption{Parameters for multilevel preconditioner ML.}
    \label{tab:sa_setup}
  \end{center}
\end{table}

To embed the solver in the physical simulation code (\opal~\cite{opal})
we utilized the Independent Parallel Particle Layer (\ippl~\cite{ippl}).
This library is an object-oriented framework for particle based
applications in computational science designed for the use on
high-performance parallel computers.  In the context of this paper
\ippl\ is only relevant because \opal\ uses \ippl\ to represent and
interpolate the particles at grid points with a charge conserving
Cloud-in-Cell area weighting scheme.  ML requires an
\texttt{Epetra\_Map} handling the parallel decomposition to create
parallel distributed matrices and vectors. To avoid additional
communication the \texttt{Epetra\_Map} and the \ippl\ field are determined
to have the same parallel decomposition. In this special case the task
of converting the \ippl\ field to an \texttt{Epetra\_Vector} is as
simple as looping over local indices and assigning values.  

A particle based domain decomposition technique based on recursive coordinate
bisection is used (see Section~\ref{sec:physrun}) to parallelize the computation
on a distributed memory environment.  One, two and three-dimensional
decompositions are available.  For problems of beam dynamics with highly
nonuniform and time dependent particle distributions, a dynamic load balancing
is necessary to preserve the parallel efficiency of the particle integration
scheme.  Here, we rely on the fact that \ippl\ attains a good load balance of
the data.



We use the solution of one time step as the initial guess for the next
time step.


\subsection{Domains}

The simplest domains under consideration are regular, rectangular
domains.  This domains are used by the FFT solver with so-called
open-space boundary conditions as described in \cite{hoea:88}. 


Our \oursolver\ solver is not restricted to rectangular domains.  It can handle
irregular domains as the ones introduced in the next section.

\subsection*{Non-rectangular domains}

To properly handle emerging irregular domains we implemented an abstract
class providing an interface to query the discretization near the
boundary.  Every implementation of an irregular domain has to identify
boundary points and provide the stencil for \textit{near-boundary}
points given one of the extrapolation schemes discussed in
Section~\ref{sec:discr}.  Boundary points are stored in \texttt{STL}
containers. Essentially the coordinate value of a gridline is mapped to
its intersection values, providing a fast look-up table for a given
gridline.

In this work we use the \oursolver\ solver mainly for cylindrical domains with
an elliptic base area.  These domains can be characterized by means of
two parameters: the semi-major and semi-minor axis.  We compute the
intersection points of the grid with the elliptical domain boundary by
using its implicit representation and subsequently store them into a
\texttt{STL} container.  These intersections have to be recomputed
whenever the parameters of the ellipse or the mesh spacings change.

\section{Numerical Experiments and Results}
\label{sec:results}

In this section we discuss various numerical experiments and results
concerning different variants of the preconditioner and comparisons of
solvers and boundary extrapolation methods.  Unless otherwise stated the
measurements are done on a tube embedded in a rectangular equidistant
$256\times256\times256$ grid.  This is a common problem size in beam
dynamics simulations.  Most of the computations were performed on the
Cray XT4 cluster of the Swiss Supercomputing Center (CSCS) in Manno,
Switzerland. Computations up to 512~cores were conducted on the XT4
cluster (\textsc{Buin}) with 468 AMD dual core Opteron 2.6\,GHz
processors and a total of 936 GB DDR RAM on a 7.6\,GB/s interconnect
bandwith network.  Larger computations were performed on the XT3 cluster
(\textsc{Palu}) with 1692 AMD dual core Opteron 2.6\,GHz processors and a
total of 3552\,GB DDR RAM interconnected with a 6.4\,GB/s interconnect
bandwith network.

Throughout this section we will report the timings of portions of the
code as listed in Table~\ref{tbl:timings_description}.
\begin{table}[ht]
  \begin{center}
    \begin{tabular}{ll}
      \hline
      name & description \\
      \hline
      construction & time for constructing the ML hierarchy \\
      application  & time for applying the ML preconditioner \\
      total ML     & total time used by ML ($=$ construction $+$ application) \\
      solution     & time needed by the iterative solver \\
      \hline
    \end{tabular}
    \caption{Description of various timings used.}
    \label{tbl:timings_description}
  \end{center}
\end{table}

\subsection{Comparison of Extrapolation Schemes}

For validation and comparison purposes we applied our solver to a
problem with a known analytical solution.  We solved the Poisson problem
with homogeneous Dirichlet boundary conditions ($\phi = 0$) on a
cylindrical domain $\Omega = \{ |r|<\frac{1}{2} \} \times
(-\frac{1}{2},\frac{1}{2})$.  The axisymmetric charge density
\begin{equation*}
  \rho = -\left( \pi^2 r^2 - \frac{\pi^2}{4} - 4 \right)
  \sin(\pi(z-0.5))
\end{equation*}
gives rise to the potential distribution
\begin{equation*}
  \phi(r, \theta, z) = \left( \frac{1}{4}-r^2 \right) \sin(\pi(z-0.5)).
\end{equation*}
The charge density in our test problem is smoother than in real
problems.  Nevertheless, it is very small close to the boundary.  This
reflects the situation in particle accelerators where most particles are
close to the axis of rotation and, thus, charge densities are very small
close to the boundary. 

We measure the error on the grid $\Omega_h$ with mesh spacing $h$ in the
discrete norms
\begin{gather*}
  \Vert e_h \Vert_2 = \Vert \hat{\phi}_h - \phi
  \Vert_2 =  \sqrt{h^3 \sum_{i \in 
      \Omega_h} \vert (\hat{\phi}_{i,h}-\phi_i)\vert^2}, \\
  \Vert e_h \Vert_\infty = \Vert \hat{\phi}_h - \phi
  \Vert_\infty =  \max_{i \in \Omega_h} 
  \vert \hat{\phi}_{i,h} - \phi_i \vert,
\end{gather*}
where $\hat{\phi}_h$ is the approximation of the solution $\phi$ on
$\Omega_h$, and $e_h$ denotes the corresponding error.
The convergence rate is approximately
\begin{equation*}
  r = \log_2 \left( \frac{\Vert e_{2h} \Vert}{\Vert e_h
      \Vert} \right).
\end{equation*}
\begin{table*}[htb]
  \begin{center}
    \begin{tabular}{llllll}
      \hline
      $h$ & $\Vert e_h \Vert_2$ & $r$ & $\Vert e_h
      \Vert_\infty$ & $r$ & $\Vert e_h \Vert_\infty /
      \Vert \phi \Vert_\infty$ \\ [0.2ex] \hline 
      \vphantom{\rule{0pt}{2.7ex}}
      $1/64$  & $2.162 \times 10^{-3}$ & & $7.647 \times 10^{-3}$ & &
      $3.061 \times 10^{-2}$ \\ 
      $1/128$ & $1.240 \times 10^{-3}$ & 0.80 & $4.153 \times 10^{-3}$ &
      0.88 & $1.662 \times 10^{-2}$ \\ [0.2ex]
      \hline
    \end{tabular}
    \caption{Solution error for constant extrapolation, $d=3$.}
    \label{tbl:rel_error_const}
  \end{center}

  \begin{center}
    \begin{tabular}{llllll}
      \hline
      $h$ & $\Vert e_h \Vert_2$ & $r$ & $\Vert e_h
      \Vert_\infty$ & $r$ & $\Vert e_h \Vert_\infty /
      \Vert \phi \Vert_\infty$\\ [0.2ex]
      \hline 
      \vphantom{\rule{0pt}{2.7ex}}
      $1/64$  & $2.460 \times 10^{-5}$ & & $6.020 \times 10^{-5}$ & &
      $2.410 \times 10^{-4}$ \\
      $1/128$ & $6.226 \times 10^{-6}$ & 1.98 & $1.437 \times 10^{-5}$ &
      2.07 & $5.751 \times 10^{-5}$ \\ [0.2ex]
      \hline
    \end{tabular}
    \caption{Solution error for linear extrapolation, $d=3$.}
    \label{tbl:rel_error_lin}
  \end{center}

  \begin{center}
    \begin{tabular}{llllll}
      \hline
      $h$ & $\Vert e_h \Vert_2$ & $r$ & $\Vert e_h
      \Vert_\infty$ & $r$ & $\Vert e_h \Vert_\infty /
      \Vert \phi \Vert_\infty$\\ [0.2ex]
      \hline 
      \vphantom{\rule{0pt}{2.7ex}}
      $1/64$  & $5.581 \times 10^{-6}$ & & $1.689 \times 10^{-5}$ & &
      $6.761 \times 10^{-5}$ \\ 
      $1/128$ & $1.384 \times 10^{-7}$ & 2.01 & $4.550 \times 10^{-6}$ &
      1.89 & $1.820 \times 10^{-5}$ \\ [0.2ex]
      \hline
    \end{tabular}
    \caption{Solution error for quadratic extrapolation, $d=3$.}
    \label{tbl:rel_error_quad}
  \end{center}
\end{table*}

We solved the Poisson equation with the boundary extrapolation methods
introduced earlier.  The errors are listed in
Tables~\ref{tbl:rel_error_const}--\ref{tbl:rel_error_quad}.  The numbers
confirm the expected convergence rates, i.e., linear for the constant
extrapolation and quadratic for the linear and quadratic
extrapolation~\cite{joma:05}.  The results obtained with the
linear extrapolation scheme are more accurate than constant
extrapolation by two orders of magnitude.  Quadratic extrapolation is
again more accurate than linear extrapolation, but for both norms by
only a factor~3 to~5.  It evidently does not make sense to use constant
extrapolation as the cost of solving with linear boundary extrapolation
is equal.  In contrast, the quadratic boundary treatment entails the
drawback that discretization matrices lose symmetry.  They are still
positive definite, however.  In the particular setting of this test
problem as well as in others we have investigated (e.g.\ in real
particle simulations) the system matrices were just `mildly'
nonsymmetric such that PCG could be applied savely and without
performance loss.


\subsection{ML variations}\label{sec:ml_var}

Multilevel preconditioners are highly sophisticated preconditioners.
Not surprisingly, their construction is very time consuming.  To build
an SA-AMG preconditioner (1) the ``grid'' hierarchy (including
aggregation and construction of tentative prolongator), (2) the final
grid transfer operators (smoothing the prolongators), and (3) the coarse
grid solver have to be set up.

In the following subsections we investigate various variants of the
preconditioner or more precisely variants of the construction of the
preconditioner when solving a sequence of related Poisson problems.

The default variant builds a new preconditioner in every time step.  In
the sequel we will investigate how costly this is.  Other variants reuse
portions of previous preconditioners.

We compare with the FFT-based Poisson solver~\cite{hoea:88} that
\textsc{OPAL} (version 1.1.5) provides for open-space boundary conditions.  
The FFT kernel is based  on a variant of the FFTPACK library~\cite{FFTPACK, Swar:82}.

\subsubsection*{Reusing the aggregation hierarchy}

Since the geometry often changes only slowly in time, the
\emph{topology} of the computational grid does not or only rarely alter.
Therefore, it makes sense to reuse the aggregation hierarchy and in
particular the tentative prolongators for some or all iterations.  Only
smoothed prolongators and other components of the hierarchy, like
smoothers and coarse solver, are recomputed~\cite[p.16]{gsht:06}.
This leads to a preconditioner variation in which the aggregation
hierarchy is kept as long as possible.  The numbers in
Table~\ref{tbl:timings_reuse_hierarchy} show that this minor change in
the set up phase reduces construction times by approximately 30\%.

\begin{table}[ht]
  \begin{center}
    \begin{tabular}{ccc}
      \hline
      \rule{15mm}{0mm} & \rule{30mm}{0mm} & \rule{30mm}{0mm} \\[-4mm]
      cores & average of 10 & one \\
      &time steps (s) &  time step (s) \\
      \hline
      16  & 6.98 & 10.3 \\
      32  & 4.36 &  6.44 \\
      64  & 2.38 &  3.48 \\
      128 & 1.34 &  1.91 \\ 
      256 & 0.735 &  1.04 \\
      512 & 0.518 &  0.745 \\
      \hline
    \end{tabular}
    \caption{Preconditioner construction times.  Left: average cost of
      10 time steps reusing the hierarchy of the first iteration step.
      Right: cost of a time step when building the whole
      preconditioner.
    }
    \label{tbl:timings_reuse_hierarchy}
  \end{center}
\end{table}

Reusing the aggregation hierarchy is a feature provided by ML.  It is
intended for use in nonlinear systems solving.  In our simulations it
reduced the time per AMG solve in a simulation run by approximately
25\%, see Table~\ref{tbl:timings_variations_overview}.

\subsubsection*{Reusing the preconditioner}

We can be more aggressive by reusing the preconditioner of the first
iteration throughout a whole simulation.  Although the iteration count
increased, the time-to-solution reduced considerably.  To counteract an
excessive increase of the iteration steps the preconditioner can be
recomputed once the number of iterations exceeds a certain threshold.
(This was not necessary in our experiments, though.)

Applying this approach to a cylinder-shaped beam pipe, a single
preconditioner could be used throughout the entire simulation without
any severe impact on the number of iteration steps.

\begin{figure}[ht]
    \centering
    \includegraphics[width=0.8\textwidth]{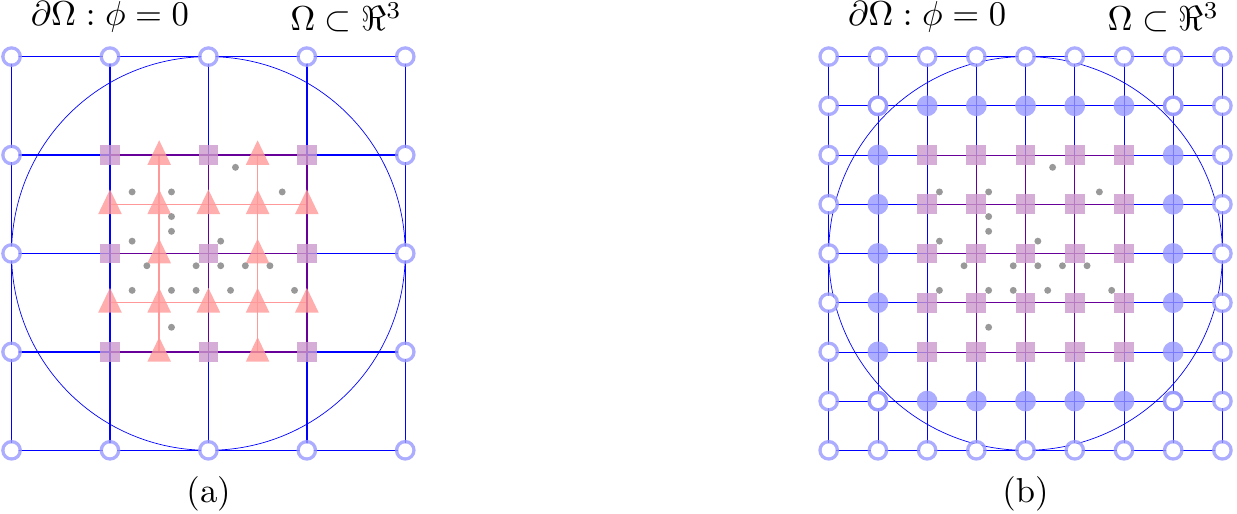}
    \centering
  \caption{Sketch of the test cases with equal number of mesh points
    (a), and equal mesh resolution (b), respectively.  Displayed are the
    shared (square), FFT only (triangle), and AMG only (filled circle) mesh
    points on a cross section of the grid plane.  Illustrative particles
    (gray) inside the FFT domain denote the charge density.}

  \label{fig:meshcmp}
\end{figure}

In the following we compare accuracy and performance of the
SA-AMG preconditioned conjugate gradient algorithm with a FFT-based
solver.  The principle difficulty in the comparison stems from the fact
that the CG-based and the FFT-based solvers use different
discretizations of the tube-shaped domain~$\Omega$.  Although both use
finite-difference discretizations, with SA-AMG the whole domain is
discretized while with FFT just a rectangular domain along the center
line of the cylinder is taken into account.  In some applications
(beam pipes of light sources) the rectangular domain is contained in $\Omega$.  In
our application the rectangular domain has a similar volume such that
the boundaries can be intertwined.  Therefore, we expect that the more
accurate boundary treatment in the iterative solver has a noticeable
positive effect on the solution.

So, we came up with two test cases as illustrated in
Figure~\ref{fig:meshcmp}.
\begin{itemize}
\item The first test case (a) displayed on the left corresponds to a
  situation where both methods have about the same number of unknowns.
  In the FFT-based approach only the close vicinity of the particle
  bunch is discretized.  In contrast, in the PCG-AMG approach the grid
  extends to the whole domain, entailing a coarser grid than with the
  FFT-based approach.

\item The second test case (b) displayed on the right corresponds to a
  situation where both methods have the same mesh resolution in the
  vicinity of the particles.  This results in a higher number of mesh
  points for the PCG-AMG approach.
\end{itemize}


We consider a cylindrical tube of radius $r=0.001$\,m.  The FFT-based
solver used a grid with $128^2\cdot256=4,194,304$ nodes.  The grid with
a similar number of points but coarser resolution consisted of 3,236,864
grid points.  To obtained the same resolution with the \oursolver\
solver, 5,462,016 grid points were required.  They are embedded in a
$166\times166\times256$ grid that coincides in the middle of the region
of simulation with the $128\times128\times256$ grid for the FFT-based
solver.  The boundary conditions were implemented by linear
extrapolation~\eqref{eq:lin_extrapol}.  In
Table~\ref{tbl:timings_variations_overview} we give execution times for
the first and second iteration in a typical simulation.  Since we start
with quite a good vector the number of steps in the second iteration is
about 30\% smaller than in the first iteration.  This accounts for the
reduced execution time in the runs where the complete preconditioners
are recomputed for each iteration.  The savings from this
straightforward mode to the two cases reusing either hierarchy or the
entire preconditioner amounts to approximately $20\%$ and $40\%$
respectively.

\begin{table}[ht]
  \begin{center}
    \begin{tabular}{cccccc}
    \hline
        solver & reusing & mesh size & mesh points & first [s] & second [s] \\
        \hline
        FFT & --- & $128\times128\times256$ & 4,194,304 & 12.3 & --- \\
        \hline
        AMG & --- & $128\times128\times256$ & 3,236,864 & 49.9 & 42.2 \\
        AMG & hierarchy & $128\times128\times256$ & 3,236,864 & --- & 35.5 \\
        AMG & preconditioner & $128\times128\times256$ & 3,236,864 & --- & 28.2 \\
        \hline
        AMG & --- & $166\times166\times256$ & 5,462,016 & 81.8 & 71.2 \\
        AMG & hierarchy & $166\times166\times256$ & 5,462,016 & --- & 60.4 \\
        AMG & preconditioner & $166\times166\times256$ & 5,462,016 & --- & 43.8 \\
        \hline
  \end{tabular}
  \caption{Simulation timings of one solve in the first and second time
    step, respectively, with $r=0.001$\,m.  Equal mesh points (above)
    and equal mesh spacings (below) for FFT and AMG.}
  \label{tbl:timings_variations_overview} \end{center}
\end{table}


We inspected the \oursolver\ results for the two mesh sizes and found no
physically relevant differences. This behaviour depends on the ratio of
bunch and boundary radius.  Based on the differences of the FFT and
\oursolver\ solver (e.g. boundary treatment and implementation details)
it is hard to come up with a `fair' comparison.  There certainly exists
a correlation between the number of performed (CG) iterations and the
time to solution.  Determining the right stopping criteria and tolerance
therefore has an important impact on the performance.  While still
achieving the same accuracy of the physics of a simulation it could be
possible to execute fewer CG iterations by using a higher tolerance.
For the measurements in Table \ref{tbl:timings_variations_overview} we
used the stopping criterion
\begin{equation*}
  \Vert r \Vert_2 \le \varepsilon  {\Vert b \Vert_2},
\end{equation*}
with the tolerance $\varepsilon = 10^{-6}$.


These results illustrate that an increase in solution accuracy of approximately
2.3 in the best case (when the domain has irregularities) is incurred in moving
from an FFT-based scheme to a more versatile approach. Of course, this more
versatile approach gives rise to increased accuracy.

\subsubsection*{Coarse level solver}

Another decisive portion of the AMG preconditioner is the coarse level
solver.  We applied either a direct solver (KLU) or used a couple of
steps of Gauss-Seidel iteration.  
Our experiments indicate that for our problems ML setup time and scalability are
not affected significantly by the two coarse level solvers. The difference in
setup and application of KLU and Gauss-Seidel  differ only within a few
percent. Only the construction of the preconditioner is cheaper with the
iterative coarse level solver.  

As stated by Tuminaro \& Tong~\cite{tuto:00} the number of iterations
done by the iterative coarse level solver is crucial for the performance of the
preconditioner.  Too many iteration steps slow down the preconditioner without a
corresponding increase of its quality.  Too
few iteration steps with the coarse level solver degrade the quality of
the overall preconditioner and lead to an increased number of steps of
the global system solver.  We tuned the iterative coarse level solver
such that the overall quality of the preconditioner was about the same
as with the direct coarse level solver.  It turned out that 3~iteration
steps sufficed.



\subsubsection*{Coarse level size}

We also tested the performance of the solver for varying sizes of the
coarsest level.  ML seems to perform best when the coarsest grid size is
around 1000.  With a limit of 1000, the coarsest grid sizes ranged from
128 to 849 when running a tube embedded in a $256 \times 256 \times 256$
grid on 16 to 512 cores.

At the same time we tried to set the size of the coarsest level
proportional to the total available number of cores in order to get a
sufficient large coarse level size.  It turned out to be very difficult
to set a coarse level size with heuristics like this.  The factorization
time increased to up to 2\,s in contrast to $0.25$\,s for the case where
the coarsest level size is limited by~1000.

\subsection{Speedup and efficiency}

\begin{figure}[htb] 
  \begin{center}
    \includegraphics[width=0.45\textwidth]{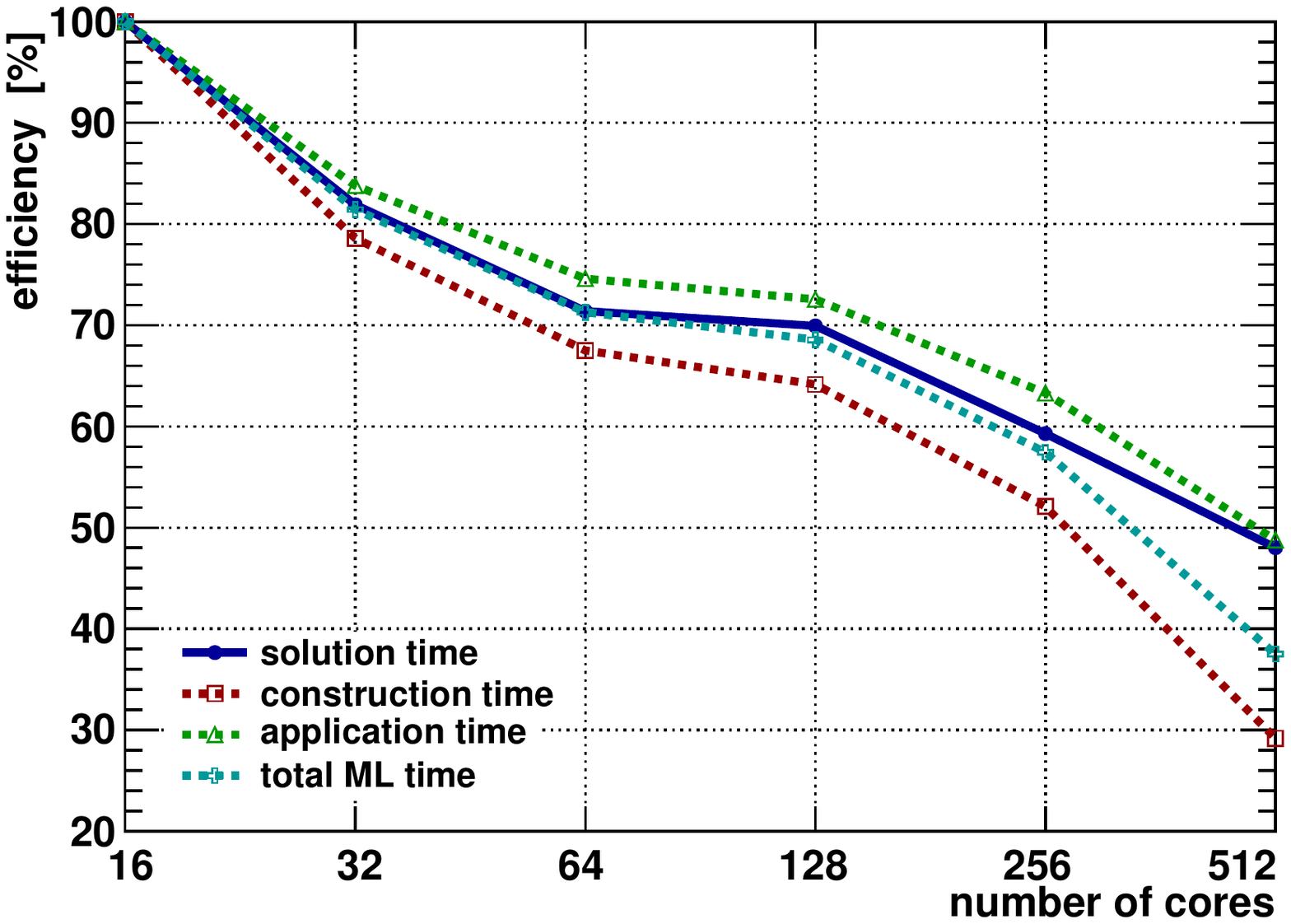}
    \includegraphics[width=0.45\textwidth]{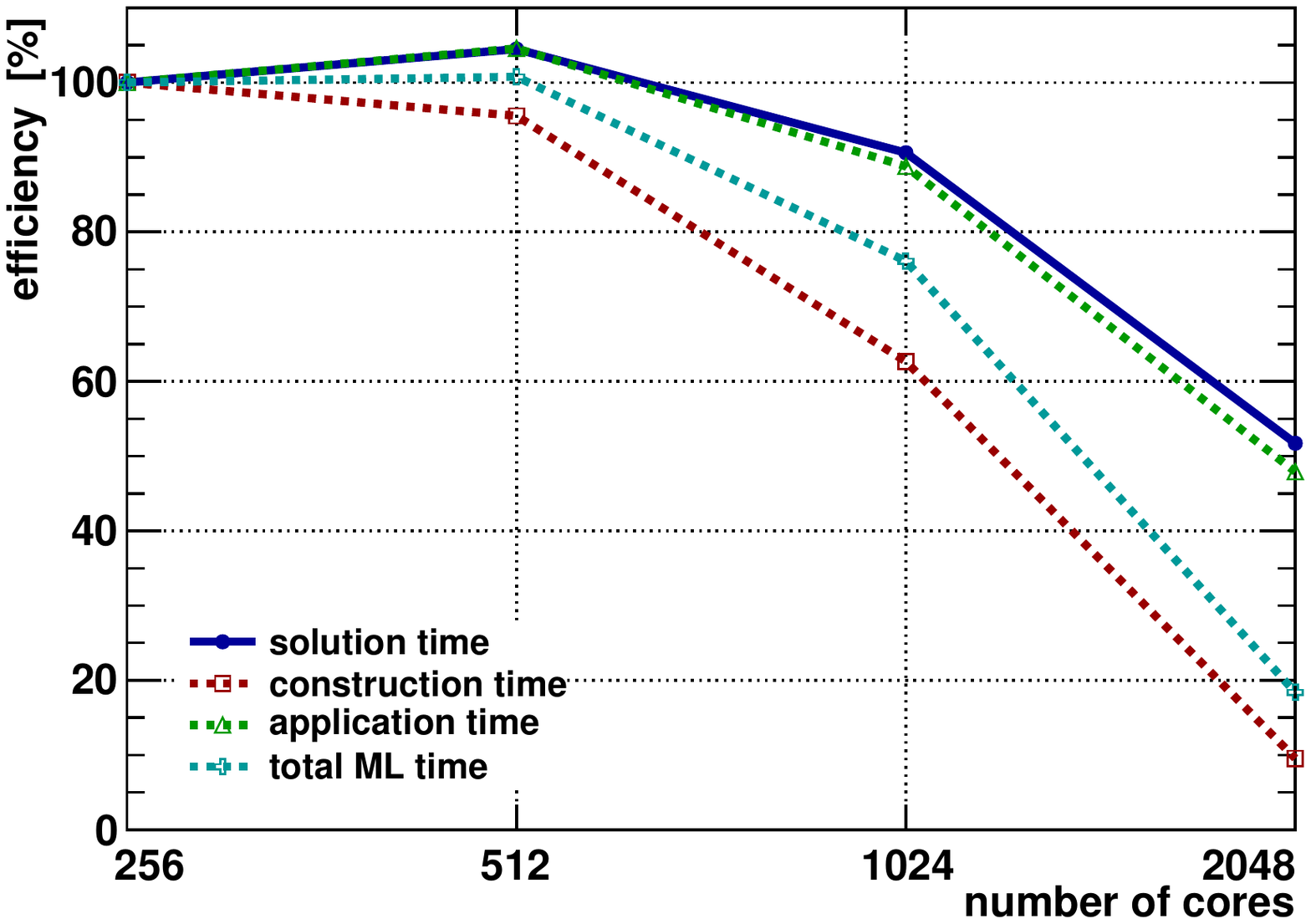}
    \caption{Efficiency for AMG on a tube embedded in a
      $256\times256\times256$ grid (left) and in a
      $512\times512\times512$ grid (right) with constant extrapolation
      at the boundary.}
    \label{fig:speedup}
  \end{center} 
\end{figure}

The efficiency of the AMG solver using linear boundary extrapolation is
shown in Figure~\ref{fig:speedup}.  On the left of this figure, the
results listed in Table~\ref{tbl:timings_solver_256} for a $256^3$ grid are
plotted.  We observe an efficiency of approximately 62\% for 256 cores
relative to the timing on 16~processors, the minimal number of
processors to solve the problem.  The efficiency dropped just below 50\%
for 512 cores.  The parallel efficiency is affected most by the poor
scalability of the construction phase of ML.  After studying various ML
timings we could not identify a specific reason that causes this loss in
efficiency for the construction phase other than the assumption that the
problem is too small with respect to the number of cores.

\begin{table}[htb]
  \begin{center}
    \begin{tabular}{ccccc}
      \hline
      cores & solution [s] & construction [s] & application [s] & total ML [s] \\
      \hline
      16  & 10.91 &  7.75 &  9.52 & 17.28 \\
      32  &  6.66 &  4.93 &  5.68 & 10.61 \\
      64  &  3.82 &  2.87 &  3.19 &  6.06 \\
      128 &  1.95 &  1.51 &  1.64 &  3.15 \\
      256 &  1.15 &  0.93 &  0.94 &  1.88 \\
      512 &  0.71 &  0.83 &  0.61 &  1.44 \\
      \hline
    \end{tabular}
    \caption{Timings for AMG on a tube embedded in a $256\times256\times256$
      grid with linear extrapolation at the boundary.}
    \label{tbl:timings_solver_256}
    \end{center}
\end{table}

\begin{table}[htb]
  \begin{center}
    \begin{tabular}{ccccc}
      \hline
      cores & solution [s] & construction [s] & application [s] & total ML [s] \\
      \hline
      256  &  10.51 &  5.79 &  8.70 &  14.49 \\
      512  &  5.03 &  3.03 &  4.16 &  7.19 \\
      1024 &  2.90 &  2.31 &  2.45 &  4.76 \\
      2048 &  2.54 &  7.58 &  2.27 &  9.85 \\
      \hline
    \end{tabular}
    \caption{Timings for AMG on a tube embedded in a $512\times512\times512$
      grid with constant extrapolation at the boundary.}
    \label{tbl:timings_solver_512} 
  \end{center}
\end{table}

Similar conclusions can be drawn for the parallel efficiency of our
solver on a $512^3$ grid with constant boundary treatment, see
Table~\ref{tbl:timings_solver_512} and Figure~\ref{fig:speedup}.  Again the
construction phase of ML scales poorly for larger numbers of cores.

Notice that by applying the improvements discussed in
Section~\ref{sec:ml_var}, i.e., reusing (parts of) the preconditioner,
the time needed for the construction phase can be reduced significantly
or avoided altogether.  If the preconditioner has to be built just once
in an entire simulation the efficiency will get close to the 52\% that
we measured for the solution phase.

\begin{figure}[htb] 
  \begin{center}
    \includegraphics[width=0.5\textwidth]{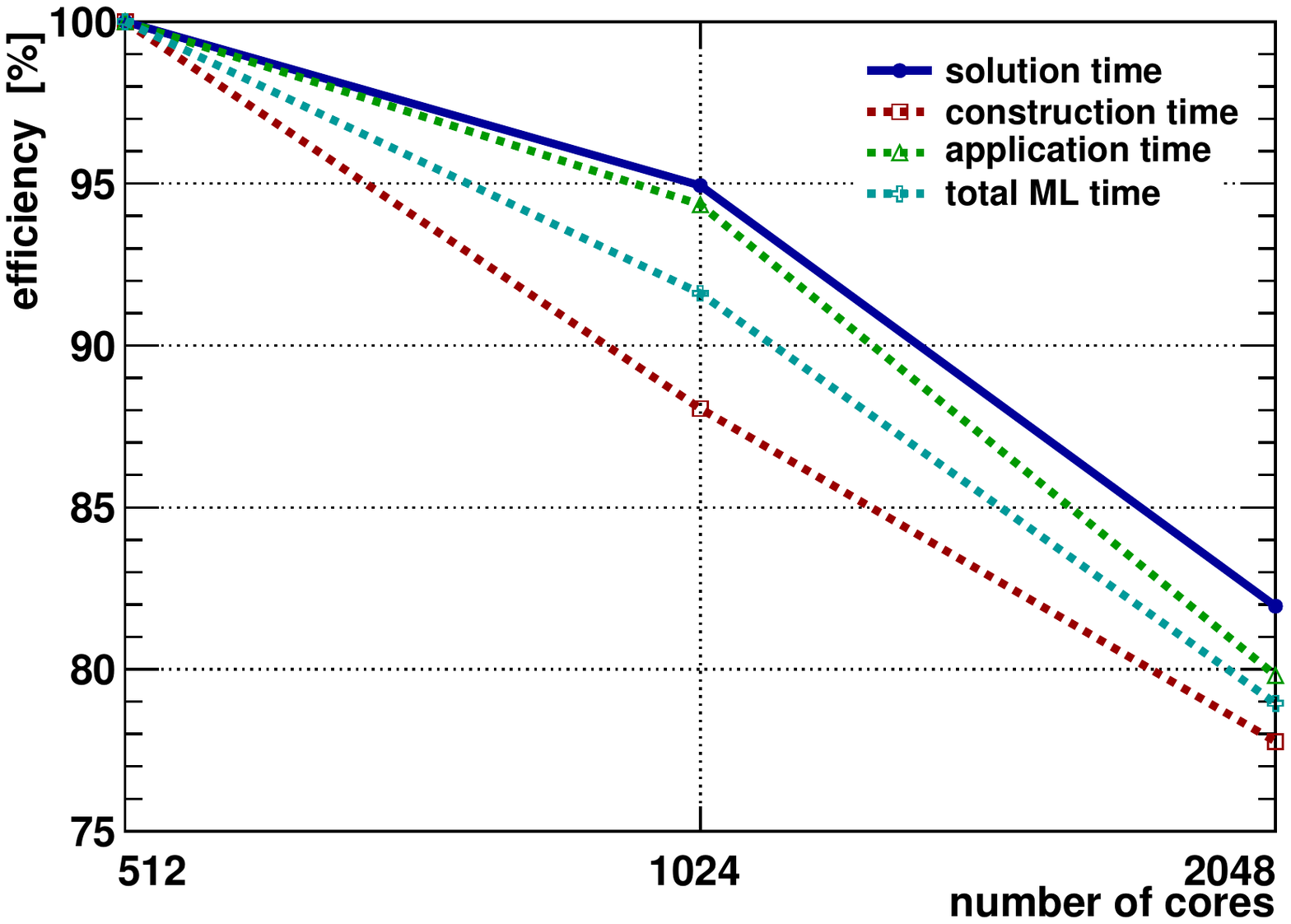}
    \caption{Efficiency for AMG on a tube embedded in a
      $1024\times1024\times1024$ grid with linear extrapolation at the
      boundary, corresponding to the numbers in
      Table~\ref{tbl:timings_solver_1024}.}
    \label{fig:speedup_1024}
  \end{center} 
\end{figure}

Finally, in Table~\ref{tbl:timings_solver_1024} we report on timings
obtained for the tube embedded in a $1024\times1024\times1024$ grid.  In
Figure~\ref{fig:speedup_1024} the corresponding efficiencies are listed.
For this large problem we observe good efficiencies.  The solver runs at
82\% efficiency with 2048 cores relative to the 512-cores performance.
The construction phase is still performing the worst with an efficiency
of 73\%.  In this setup the problem size is still reasonably large when
employing 2048 cores.  This consolidates our understanding of the
influence of the problem size on the low performance of the aggregation
in ML.

\begin{table}[htb]
  \begin{center}
    \begin{tabular}{ccccc}
      \hline
      cores & solution [s] & construction [s] & application [s] & total ML [s] \\
      \hline
      512  &  35.83 &  20.78 &  29.53 &  50.31  \\
      1024 &  18.87 &  11.80 &  15.65 &  27.46  \\
      2048 &  10.93 &  6.68 &  9.25 &  15.93   \\
      \hline
    \end{tabular}
    \caption{Timings for AMG on a tube embedded in a
      $1024\times1024\times1024$ grid with linear extrapolation at the
      boundary.}
    \label{tbl:timings_solver_1024}
  \end{center}
\end{table}
Tables~\ref{tbl:timings_solver_256}--\ref{tbl:timings_solver_1024}
provide a few data to investigate weak scability.  The timings for 2048
processors in Table~\ref{tbl:timings_solver_1024} should ideally equal
those for 256 processors in Table~\ref{tbl:timings_solver_512}.  In
fact, they are quite close.  A comparison of the timings in
Tables~\ref{tbl:timings_solver_512} and~\ref{tbl:timings_solver_256} is
not so favorable.  The efficiency is at most 84\%.  The numbers for 2048
processors in Table~\ref{tbl:timings_solver_512} show that the
construction of the multilevel preconditioner becomes excessively
expensive if the number of processors is high.

\subsection{Open-space vs.\ PEC Boundary Conditions} 
\label{sec:physrun}

In this section we compare the impact of two different boundary
conditions in the setting of a physical simulation consisting of an
electron source (4 MeV) followed by a beam transport section
\cite{schiet:08}.  As the pipe radius gets close to the particles in the
beam, the fields become nonlinear due to the image charge on the pipe.
We compare the root-mean-square (rms) beam size in a field free region
(drift) of a convergent beam, cf.~\cite[pp.171ff]{wied:07}.  The
beam pipe radius is $r = 0.00085$\,m in case of the PGC-MG solver.  This
is an extreme case in which the particles fill almost the whole beam
pipe and hence the effect is very visible. 
The capability to have an exact representation of
the fields near the boundaries is very important, because the beam pipe radius is an important optimization quantity,
towards lower construction and operational costs in the design and operation of future particle accelerators. 

\begin{figure}[ht]
  \begin{center}
    \includegraphics[width=0.5\textwidth]{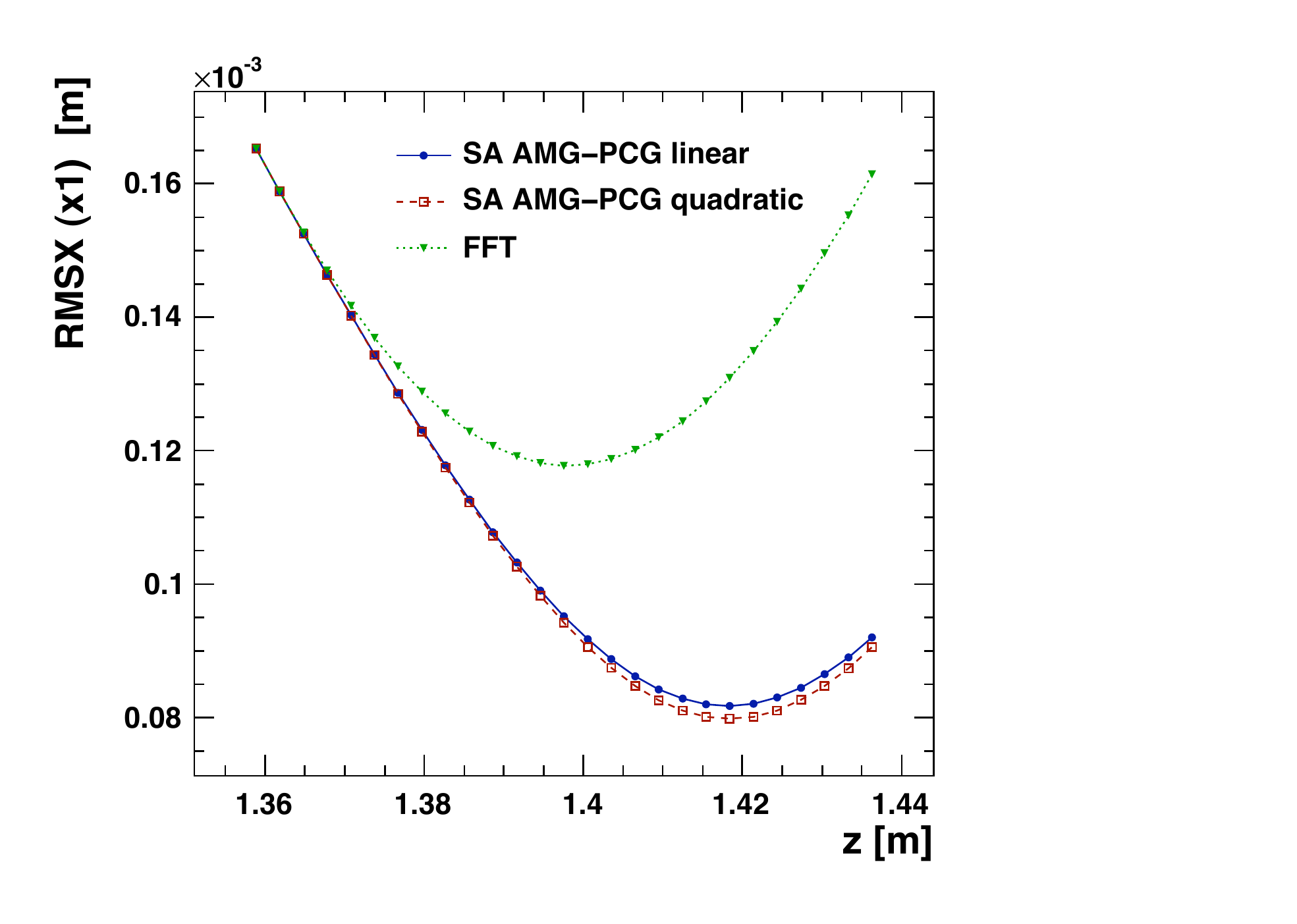}
    \caption{Comparison of rms beam size vs.\ position of a convergent
      beam. The FFT solver is applied for open-space boundary
      conditions, for PEC two variants of the SA AMG-PCG solver is used:
      with linear and quadratic boundary interpolation.  (region of
      interest magnified). The computational domain $\Omega$ is a
      cylinder with $r=0.00085$\,m \label{fig:vareps}.}
  \end{center} 
\end{figure}

In Figure~\ref{fig:vareps} we compare rms beam sizes for the two
boundary conditions applied to the boundary of a cylinder with
elliptical base-area as described in Section~\ref{sec:discr}.


The differences are up to $~40\%$ (at $z=1.435$\,m) in rms beam size, when comparing the PEC and the open-space approach. We clearly see
the shift of the beam size minimum (waste) towards larger $z$ values and
a smaller minima, which means that the force of the self fields are
larger when considering the beam pipe.
This increase in accuracy justifies the 
accurate boundary treatment, in situations where the spatial extent of
the beam is comparable with that of the beam pipe.

\section{Conclusion}
\label{sec:concl}

We have presented a scalable Poisson solver suitable to handle domains
with irregular boundaries as they arise, for example, in beam dynamics
simulations.  The solver employs the conjugate gradient algorithm
preconditioned by smoothed aggregation based AMG (SAAMG-PCG).  PEC and
open-space boundary approximations have been discussed.  A real world
example where the solver was used in a beam dynamics code (OPAL) shows
the relevance of this approach by observing up to $~40\%$ difference in
the RMS beam size when comparing to the FFT-based solver with open
domains.  The code exhibits excellent scalability up to 2048 processors
with cylindrial tubes embedded in meshes with up to $1024^3$ grid
points.  In the very near future, this approach will enable precise beam
dynamics simulations in large particle accelerator structures with a
level of detail not obtained before.

In real particle simulations (and other test cases encountered) system
matrices arising from quadratic boundary treatment are only `mildly'
nonsymmetric such that PCG can be applied.

Planed future work includes adaptive mesh refinement in order to reduce
the number of grid points in regions that are less relevant for the
space charge calculation.  The boundary treatment for simply connected
geometries will be extended to cope with more realistic geometries.  This
new method, designed for accurate 3D space charge calculations, will be
used in the beam dynamics simulations for the SwissFEL project, a next
generation light source foreseen to be built in Switzerland.

\section*{Acknowledgments}

The majority of computations have been performed on the Cray XT4 in the
framework of the PSI CSCS ``Horizon'' collaboration.  We acknowledge the
help of the XT4 support team at CSCS, most notably Timothy Stitt.
The XT5 performance results have been obtained on a Cray XT5
supercomputer system, courtesy of Cray Inc.  We acknowledge the help of
Stefan Andersson (Cray Inc.).
We thank Jonathan Hu and Raymond Tuminaro (Sandia National Laboratories)
for valuable information and help regarding the ML package.
Special thanks are addressed to reviewers and editors for their valuable
suggestions.

\bibliography{paper}
\bibliographystyle{elsarticle-num}

\end{document}